\documentclass{aa}
\usepackage{txfonts}
\usepackage{graphicx}
\usepackage{natbib}
\bibpunct{(}{)}{;}{a}{}{,}

\begin{document}

\title{Low Power Compact Radio Galaxies at High Angular Resolution}

\author{ M. Giroletti\inst{1,2} 
    \and G. Giovannini\inst{1,2} 
    \and G. B. Taylor\inst{3,4} }
\institute{Dipartimento di Astronomia, Universit\`a di Bologna, via Ranzani 1, 40127 Bologna, Italy
     \and Istituto di Radioastronomia, via Gobetti 101, 40129, Bologna, Italy
     \and Kavli Institute of Particle Astrophysics and Cosmology, Menlo Park, CA 94025, USA
     \and National Radio Astronomy Observatory, P.O. Box O, Socorro, NM 87801}

\date{Received / Accepted }

\abstract{ We present sub-arcsecond resolution multi-frequency (8 and 22 GHz)
  VLA images of five low power compact (LPC) radio sources, and phase
  referenced VLBA images at 1.6 GHz of their nuclear regions. At the VLA
  resolution we resolve the structure and identify component positions and flux
  densities. The phase referenced VLBA data at 1.6 GHz reveals flat-spectrum,
  compact cores (down to a few milliJansky) in four of the five sources. The
  absolute astrometry provided by the phase referencing allows us to identify
  the center of activity on the VLA images. Moreover, these data reveal rich
  structures, including two-sided jets and secondary components.  On the basis
  of the arcsecond scale structures and of the nuclear properties, we rule out
  the presence of strong relativistic effects in our LPCs, which must be
  intrinsically small (deprojected linear sizes $\la 10$ kpc). Fits of
  continuous injection models reveal break frequencies in the GHz domain, and
  ages in the range $10^5-10^7$ yrs. In LPCs, the outermost edge may be
  advancing more slowly than in more powerful sources or could even be
  stationary; some LPCs might also have ceased their activity. In general, the
  properties of LPCs can be related to a number of reasons, including, but not
  limited to: youth, frustration, low kinematic power jets, and short-lived
  activity in the radio.

\keywords{Radio continuum: galaxies -- Galaxies: active -- Galaxies:
    jets -- Galaxies: individual: 0222+36, 0258+35, 0648+27, 1037+30,
    1855+37} }

\maketitle

\email{giroletti@ira.cnr.it, ggiovann@ira.cnr.it, gtaylor@nrao.edu}

\section{Introduction}

Radio sources come in different well known and understood morphologies.
Extended (kiloparsec scale) radio galaxies are divided between FR I and FR II
radio galaxies, according to their morphology and radio power
\citep{fan74}. \citet{led96} showed that the separation between these two
classes is related to both the total radio power and the optical magnitude of
the host galaxy. \citet{ghi01} showed how this separation can be interpreted in
terms of accretion rates below or above a critical upper limit for maintaining
an optically thin advection dominated accretion flow \citep[ADAF,][]{nar95}
regime.

Compact (sub-kiloparsec) radio sources are also divided on the basis of
different morphologies.  Some of them appear as compact radio sources because
of projection effects; this includes BL Lacertae objects \citep[e.g., Mkn
501,][]{gir04} and flat spectrum radio quasars (FSRQ), i.e., sources oriented
at small angles with respect to the line of sight. Their compactness is due to
projection effects, and to strong relativistic effects which give rise to
one-sidedness, superluminal motions, and high brightness temperatures.

Other high power radio sources appear intrinsically small and are not affected
by relativistic effects, since we do not see beamed relativistic jets but
simply the regions of interaction between jets and the ISM (hot spots), which
are advancing at velocities of the order of 0.2c \citep{ows98,pol02,gir03}.
The Compact Symmetric Objects class \citep[CSOs, see e.g.][]{wil94,gug05} is a
good example of this population: it is composed of small sources ($<1$ kpc),
with emission on both sides of the central engine.  On the basis of kinematics
as well as spectral arguments, these objects are interpreted as young radio
galaxies with ages $\la 10^4$ years and are expected to evolve into kiloparsec
scale radio galaxies \citep{ows98,mur99,gir03}.  The number of these sources
appears, however, to be too high with respect to the general population of
radio galaxies \citep{fan90,ode97}, therefore it has been suggested that their
radio power decreases and/or expansion slows with time to account for the
observed population of giant radio galaxies \citep{kai97,blu99,fan03,lar04}.

In catalogues of radio sources selected at low frequency there is another class
of radio sources whose properties are not yet well known. We have named these
low power compact (LPC) radio sources. Most of these sources do not have a flat
radio spectra and show a moderately steep spectral index. Moreover, their host
galaxies do not show signatures of strong nuclear activity in the optical and
X-ray bands. X-ray luminosities, or upper limits, are typically of the order
10$^{40}$ erg s$^{-1}$ \citep[e.g.][]{can99}. In the optical, central compact
cores are seldom visible; and even when they are detected, these cores show a
similar nature to the synchrotron optical cores of FR I radio galaxies rather
than signatures of thermal emission from an efficiently radiating accretion
disk \citep{cap02}. The radio powers are typically below 10$^{25}$ W Hz$^{-1}$
at 1.4~GHz. Little variability is present in the radio fluxes, although this
can be accounted for by the paucity of measurements.

The small size of LPCs could be directly related to the low radio power: the
central AGN has insufficient power to drive the relativistic jet out of the
dense ISM present in the central regions of the host galaxy.  This could be the
case for sources like NGC 4278 \citep{gir05} and may constitute the link
between radio loud and radio quiet AGN such as Seyfert galaxies. However, most
of these low power compact radio sources have an intrinsic radio power in the
same range as that of low power giant FR I radio galaxies. In this case, the
reason for the compactness of these sources is unclear. It could be again any
of the previous physical reasons: geometrical-relativistic effects, low age, or
frustration by a denser than average ISM.  It could be that all of these
effects are present, and in addition some sources may even be prematurely dying
\citep{mar03,kun04,gug05}.

All these possibilities are interesting: in the case of geometric-relativistic
effects, we would have a population of low power BL-Lacs where relativistic
jets are present in spite of the low power of the AGN; in the case of young
radio sources, we could test if their number is in agreement with current
populations and discuss whether they will become giant radio galaxies or not.
The LPC sources could be related to CSOs, but at lower radio power with respect
to most CSOs studied up to now. Finally, frustration has been excluded as an
explanation to the origin of CSOs, but could apply in some LPC sources, so it
is important to measure kinematic ages for symmetric LPC sources.

To investigate these points we have selected from the Bologna Complete Sample
\citep[BCS,][]{gio05} compact ($<$ 10 kpc) radio sources with a total low
frequency radio power $P_\mathrm{408\,MHz} < 10^{25.5}$ W Hz$^{-1}$ whose
sub-parsec structure was not well defined. We are studying their properties
with Very Long Baseline Array (VLBA) and high frequency, high resolution Very
Large Array (VLA) observations.  Two sources fulfilling the same requirements
(the CSO 4C\,31.04 and NGC\,4278) have been already discussed in published
papers \citep{gir03,gir05}. Here we present new results for 5 radio galaxies:
\object{0222+36}, \object{0258+35}, \object{0648+27}, \object{1037+30}, and
\object{1855+37} (see Table~\ref{table1}). The remaining compact sources will
be observed in the near future and discussed in light of the statistical
properties of the BCS.

The paper is laid out as follows: in \S 2 we give the details of our
observations, in \S 3 we present the results about single sources, and in \S 4
we discuss the properties of our subsample with respect to other classes of
compact sources; we present our main conclusions in \S 5. Throughout this
paper, we make use of H$_0$ = 70 km sec$^{-1}$ Mpc$^{-1}$, $\Omega_M$ = 0.3 and
$\Omega_\Lambda$ = 0.7. Spectral indices are defined such that $S(\nu) \propto
\nu^{-\alpha}$.

\begin{table}
\caption{Sources Basic Parameters\label{table1}}
\begin{tabular}{cccccc}
\hline\hline
Source & $z$ & Log $P_\mathrm{408\,MHz}$ & $M_V$ \\
 & & (W Hz$^{-1}$) & (mag) \\
\hline
0222+36 & 0.0334 & 23.91 & $-22.3$  \\
0258+35 & 0.0165 & 24.37 & $-21.8$  \\
0648+27 & 0.0414 & 24.02 & $-23.3$  \\
1037+30 & 0.0911 & 25.37 & $-21.7$  \\
1855+37 & 0.0552 & 24.65 & $-23.8$  \\
\hline \\
\end{tabular}
\end{table}

\section{Observations and Data Reduction}

\subsection{VLA observations}

\begin{table*}
\caption{Log of VLA Observations\label{log_vla}}
\begin{tabular}{cccccccc}
\hline\hline
 & absolute &   & \multicolumn{2}{c}{8.4 GHz} & \multicolumn{3}{c}{22 GHz} \\
 & flux density &   & minutes & phase and amp. & minutes & amplitude & phase \\
date & calibrator & target & on source & calibrator & on source & calibrator &  calibrator \\
\hline
28/6 & 3C\,147 & 0648+27 & 11 & 0645+213 & 18 & 0741+312 & 0657+243 \\
     & 3C\,48  & 0222+36 & 11 & 0237+288 & 18 & 0237+288 & 0230+405 \\
     &         & 0258+35 & 11 & 0237+288 & 18 & 0237+288 & 0310+382 \\
29/6 & 3C\,286 & 1037+30 & 9.5 & 0956+252 & 12 & 0956+252 & 1037+285 \\
 3/7 & 3C\,286 & 1855+37 & 10.5 & 1800+388 & 18 & 1800+388 & 1912+376 \\
\hline
\end{tabular}
\end{table*}

\begin{table*}
\caption{Observational parameters\label{obs_para}}
\begin{tabular}{lclrclrclr}
\hline\hline
 & \multicolumn{3}{c}{VLA @8.4 GHz} & \multicolumn{3}{c}{VLA @22 GHz} & \multicolumn{3}{c}{VLBA @1.6 GHz} \\
Name & beam & noise & peak & beam & noise & peak & beam & noise & peak \\
     & ($\arcsec \times \arcsec,\, ^\circ$) & \multicolumn{2}{c}{(mJy beam$^{-1}$)} & ($\arcsec \times \arcsec, ^\circ$) & \multicolumn{2}{c}{(mJy beam$^{-1}$)} & (mas $\times$ mas, $^\circ$) & \multicolumn{2}{c}{(mJy beam$^{-1}$)} \\ 
\hline
0222+36 & $0.23 \times 0.19, -52$ & 0.19 &  164 & $0.11 \times 0.10, 89$  & 0.17 &  89 & $10.5 \times 4.4, 9 $ & 0.33 & 61.2 \\
0258+35 & $0.22 \times 0.19, -39$ & 0.21 &  126 & $0.12 \times 0.12, 0$   & 0.19 &  56 & $10.4 \times 4.3, 6 $ & 0.40 &  7.5 \\
0648+27 & $0.27 \times 0.24, 65$  & 0.16 & 11.5 & $0.12 \times 0.10, -76$ & 0.19 & 2.3 & $10.6 \times 4.4, 13$ & 0.57 &  4.4 \\
1037+30 & $0.21 \times 0.20, 9$   & 0.09 & 15.9 & $0.08 \times 0.08, 15$  & 0.37 & 2.1 & $10.2 \times 5.1, 22$ & 0.55 &  2.3 \\
1855+37 & $0.27 \times 0.26, 83$  & 0.06 & 0.62 & $0.19 \times 0.17, -88$ & 0.18 & -- & $11.9 \times 5.3, -47$ & 0.38 & -- \\
\hline
\\
\end{tabular}

NOTE -- Observational parameters for VLA and VLBA images in Figs.~\ref{fig0222+36}, \ref{fig0258+35}, \ref{fig0648+27}, \ref{fig1037+30}, \ref{fig1855+37}: beam size, lowest contour (set equal to the $3\sigma$ noise levels) and peak flux densities. 1855+37 is not detected with the VLA at 22 GHz nor at 1.6 GHz with the VLBA.
\end{table*}

\begin{table*}
\caption{VLBA Observations: Source and Calibrators List\label{log_vlba}}
\begin{tabular}{lcllcc}
\hline\hline
Source & \multicolumn{2}{c}{Core Absolute Position} & Calibrator & Separation \\
 & (RA, $^{h\,m\,s}$) & (Dec, $^{\circ}\,'\,''$) & & ($^{\circ}$) \\
\hline
0222+36 & 02 25 27.327 &37 10 27.746 & J0226+3421 &  2.8 \\
0258+35 & 03 01 42.329 &35 12 20.298 & J0310+3814 &  3.8 \\
0648+27 & 06 52 02.517 &27 27 39.405 & J0646+3041 &  3.5 \\
1037+30 & 10 40 29.945 &29 57 57.791 & J1037+2834 &  1.5 \\
1855+37 & 18 57 37.52  &38 00 33.7   & J1912+3740 &  3.7 \\
\hline \\
\end{tabular}

NOTE -- Incomplete coordinates for 1855+37 are due to the lack of a detection
in the VLBA and VLA 22 GHz data. Therefore, the absolute position reported here
is taken from the VLA data at 8.4 GHz.
\end{table*}

VLA observations of five objects (0222+36, 0258+35, 0648+27, 1037+30, and
1855+37) were obtained in three observing runs in 2003 June 28 and 29 and 2003
July 10. The array was in ``A'' configuration (maximum baseline 35.4 km) and
the observing frequencies were 8.4 and 22 GHz. Standard observing schedules for
high frequency observations were prepared, including scans to determine the
primary reference pointing, and using a short (3 s) integration time and fast
switching mode (180 s on source, 60 s on calibrator) for K band (22 GHz)
scans. Primary, amplitude, and phase calibrators for each run and for each
source are given in Table~\ref{log_vla}.  Post-correlation processing and
imaging were performed with the NRAO Astronomical Image Processing System
(AIPS).  Parameters of different images are reported in Table \ref{obs_para}.

\subsection{VLBA observations}

The 5 sources are also part of a sample of 15 objects that have been observed
in phase reference mode with the VLBA, in order to study the parsec scale
structure of faint radio galaxies. Observations were done in two separate runs
on 2003 August 07 (BG136A, 24 hrs for 12 sources) and 2003 August 30 (BG136B, 6
hrs for 3 sources). We discuss here data for the 5 LPC sources only; they have
all been observed in segment A of the experiment; the VLBA structure for the
other 10 sources will be discussed in a future paper.

We observed in full polarization (RCP and LCP) with two IFs (central frequency
1659.49 MHz and 1667.49 MHz). We recorded 16 channels per frequency, for a
total aggregate bit rate of 128 Mbs. Each pointing on a target source was
bracketed by a calibrator scan in a 5 minutes duty cycle (3 minutes on source,
2 minutes on the calibrator). Two groups of (typically) 11 cycles were executed
for each source at different hour angles, resulting in a total of about 66
minutes per target, with good coverage of the $(u, v)-$plane. Calibrators were
chosen from the VLBA calibrators list to be bright and close to the source; we
report in Table \ref{log_vlba} the list of the selected calibrator and its
separation for each source discussed here. Short scans on strong sources
(4C39.25, J0237+2848) were interspersed with the targets and the calibrators as
fringe finder sources.

The correlation was performed in Socorro and the initial calibrations were done
with AIPS. Scans on J0237+2848 were used to remove IF-dependent delays and
phase offsets. After applying models for the electron content of the ionosphere
as measured by the Jet Propulsion Laboratory (JPL) on the observing dates, we
performed a global fringe fitting on all calibrators and applied the solutions
to the targets with a two-point interpolation. $R-L$ delay differences were
determined and removed with the VLBACPOL procedure using 4C39.25. We then
produced images of the calibrators in order to obtain and apply more accurate
phase and gain corrections to the sources; we also determined from preliminary
maps the absolute position of the sources, which we used thereafter. We give
the coordinates of the core candidate of each source in Table~\ref{log_vlba}.

The final calibrated single-source datasets were imported into Difmap for
imaging and self-calibration. We produced images with natural and uniform
weights, and summarize in Table \ref{obs_para} the significant parameters for
the final images.

\section{Results}
\label{results}

\begin{table*}
\caption{Results of modelfit - VLA images \label{res_vla}}
\begin{tabular}{llrrrrcc}
\hline\hline
 & & \multicolumn{2}{c}{8.4 GHz} & \multicolumn{2}{c}{22.5 GHz} \\
Source & Component & peak & flux & peak & flux & $\alpha_{8.4}^{22}$ & size \\
 & & ($\frac{\mathrm{mJy}}{\mathrm{beam}}$) & (mJy) & ($\frac{\mathrm{mJy}}{\mathrm{beam}}$) & (mJy) & & \arcsec \\
\hline
0222+36 & core    & 174 & 179 &  89 &  91 & $0.69 \pm 0.02$ & 0.05 \\
        & NE lobe & 5.7 & 7.0 & 1.5 & 2.8 & $0.93 \pm 0.04$ & 0.2 \\
        & SW lobe & 3.2 & 5.7 & 0.6 & 2.2 & $0.97 \pm 0.04$ & 0.4 \\
        & total   &  -- & 192 &  -- &  95 & $0.71 \pm 0.02$ & 1.6 \\
\hline
0258+35 & core region & 119 & 167 &  52 & 14+57 & $0.87 \pm 0.02$ & 0.08 \\
        & SE jet      &  58 & 134 &  15 &    77 & $0.56 \pm 0.02$ & 1.0 \\
        & E lobe      &  21 & 197 & 4.8 &    77 & $0.95 \pm 0.02$ & 1.6 \\
        & NW lobe     &  14 &  93 & 2.6 &    37 & $0.93 \pm 0.02$ & 1.6 \\
        & total       &  -- & 610 &  -- &   255 & $0.88 \pm 0.01$ & 3.5 \\
\hline
0648+27 & N lobe &  11 & 19 & 2.2 & 2.9+2.3 & $1.31 \pm 0.03$ & 0.6 \\
        & S lobe & 8.4 & 14 & 0.9 &     4.9 & $1.06 \pm 0.03$ & 0.6 \\
        & total  &  -- & 31 &  -- &     9.3 & $1.22 \pm 0.02$ & \\
\hline
1037+30 & core      & 2.2 & 3.9 & 1.0 & 2.6 & $0.41 \pm 0.06$ & 0.04 \\
        & jet knot  & 1.2 & 2.4 &  -- &  -- &   -- & 0.4 \\
        & S lobe    & 1.2 &  10 &  -- &  -- &   -- & 1.2 \\
        & S hotspot & 2.3 & 8.6 &  -- &  -- &   -- & 0.4 \\
        & N lobe    & 5.7 &  22 &  -- &  -- &   -- & 0.8 \\
        & N hotspot &  13 &  20 & 4.3 & 12  & $0.52 \pm 0.02$ & 0.4 \\
        & total     &  -- &  63 &  -- & 14  & $1.53 \pm 0.02$ & 3 \\
\hline 
1855+37 & core  & 0.6 &  0.9 & -- & -- & -- & 0.3 \\
        & total &  -- & 35.9 & -- & -- & -- & 8 \\
\hline \\
\end{tabular}

NOTE -- Uncertainties in Col. (7) are $1\sigma$ values based on the absolute
flux calibration and noise in the images. The core region in 0258+35 and the
northern lobe in 0648+27 are resolved in two components at 22 GHz, therefore we
give two values in Col. (6) and the spectral index is spurious. See Table
\ref{res_vlba} for a more reliable estimate of the cores spectral index. The
source 1855+37 is not detected at 22 GHz.
\end{table*}

All sources are detected in our VLA data at 8.4 GHz. The source structures are
successfully resolved, allowing a study of different components. At 22 GHz, we
have high signal to noise detections for 0222+36, 0258+35, and 0648+27; in the
source 1037+30 we detect only the core and N-W hot spots; finally, 1855+37 is
completely resolved and not even a faint nuclear source was detected.  In
sources 0222+36, 0258+35, and 0648+27 the lack of short baselines in 22 GHz
images does not affect flux density measurements, therefore a spectral index
comparison between 8.4 and 22 GHz is meaningful.  We give in Table
\ref{res_vla} the main parameters for different sources and subcomponents, such
as flux densities and spectra.

\begin{table}
\caption{Results of modelfit - VLBA images\label{res_vlba}}
\begin{tabular}{llrrcc}
\hline\hline
Source & Component & peak & flux & size & $\alpha_{1.6}^{22}$ \\ 
 & & $(\frac{\mathrm{mJy}}{\mathrm{beam}})$ & (mJy) & (mas) \\ 
\hline
0222+36 & core  & 61 & 71 & 2  & $0.05 \pm 0.02$ \\ 
        & N jet &  6 & 16 & 13 \\ 
        & S jet &  3 & 26 & 29 \\ 
        & total & -- & 102 &  \\
\hline
0258+35 & core  & 3.3 & 7.4 &  6 & $-0.24 \pm 0.02$ \\
        & blob  & 7.5 & 236 & 44 & $0.54 \pm 0.02$ \\
        & total &  -- & 243 & \\
\hline
0648+27 & core  & 4.3 & 10  & 5 & $0.47 \pm 0.03$ \\
        & jet   & 1.1 & 3.4 & 5 & \\
        & total &  -- & 12.8 & \\
\hline
1037+30 & core  & 3.4 & 3.8 & 5 & $0.14 \pm 0.04$ \\
        & hot spot & 1.0 & 40 & 60 & $0.46 \pm 0.02$ \\
        & total &  -- & 14  \\
\hline
1855+37 & \multicolumn{5}{l}{N.D.} \\
\hline \\
\end{tabular}

NOTE -- The spectral index for 0222+36 is computed considering the total flux
density detected at 1.6 GHz by the VLBA, which extends over a region smaller
than the beam of the VLA at 22 GHz. Spectral index uncertainties are $1\sigma$
values.
\end{table}

The VLBA phase referenced images in total intensity are available for all
sources except 1855+37, which is not detected. On the mas scale a resolved
structure is visible in 0222+36, 0258+35, and 0648+27, while in 1037+30 we
detect a hot spot in addition to the core, $\sim 0.8\arcsec$ north-west of it.
Thanks to successful phase-referencing, our data yield accurate positional
information which unambiguously identifies the nuclear component in these
complex sources.  Results are reported in Table \ref{res_vlba}, including
spectral information confirming the core identification. The core spectral
indices are computed between 1.6 and 22 GHz; for 0222+36 only, we consider the
total flux density detected at 1.6 GHz by the VLBA, which extends over a region
smaller than the beam of the VLA at 22 GHz. For other sources, secondary
components are well separated in the VLA images (0258+35, 1037+30), or their
flux density is negligible (0648+27).  We also place upper limits on the
fractional polarization. In fact, all images in polarized intensity are purely
noise-like, with $3\sigma$ levels of approximately 0.25 mJy/beam. Since the
sources are weak, these provide weak constraints of approximately $P < 10\%$ at
the peak position (except for 0222+36, see \ref{0222+36}).

Finally, we also collected integrated flux density measurements from the NASA
Extragalactic Database (NED), as well as from radio surveys including the
Northern VLA Sky Survey (NVSS), the Westerbork Northern Sky Survey (WENSS), and
the VLA Low-frequency Sky Survey (VLSS). These data typically cover the
frequency range between 74 MHz and 5 GHz. Our new measurements at shorter
wavelengths allow us to obtain integrated spectra spanning two and a half
orders of magnitude in frequency. We show the integrated spectra for each
source in the bottom right panels of Figs.~\ref{fig0222+36}, \ref{fig0258+35},
\ref{fig0648+27}, \ref{fig1037+30}, \ref{fig1855+37}, with best-fit continuous
injection models \citep{mur99} overlaid. Detailed results for each source are
presented in the following subsections.

\subsection{0222+36}
\label{0222+36}

\begin{table}
\caption{Spectral data for 0222+36 \label{tab0222+36sp}}
\begin{tabular}{lcccc}
\hline\hline
Frequency & Total & Halo & Lobes & Core \\
(GHz) & (mJy) & (mJy) & (mJy) & (mJy) \\
\hline
0.074 (VLSS)  & 580 & -- & -- & -- \\
0.325 (WENSS) & 380 & 188$^{\mathrm{a}}$ & 191$^{\mathrm{b}}$ & 0.9$^{\mathrm{b}}$ \\
0.365 (TEXAS) & 352 & 174$^{\mathrm{a}}$ & 177$^{\mathrm{b}}$ & 1.2$^{\mathrm{b}}$ \\
0.408 (B2)    & 337 & 170$^{\mathrm{a}}$ & 165$^{\mathrm{b}}$ & 1.8$^{\mathrm{b}}$ \\
1.4 (VLA)  & 190 & 36   & 52   & 102$^{\mathrm{c}}$ \\
5.0 (VLA)  & 224 & 8    & 36   & 180 \\
8.4 (VLA)  & 192 & nd   & 26   & 166 \\
22.5 (VLA) & 96  & nd   & 6    & 90 \\
\hline \\
\end{tabular}

NOTES - (a) the flux density of the halo is computed from subtraction of the
nuclear and lobe components from the total flux density listed in Col. (2); (b)
Lobes and core flux densities at low frequency (0.325, 0.365, and 0.408 GHz)
are extrapolated from data between 1.4 and 22 GHz (see text); (c) from VLBA
data at 1.6 GHz
\end{table}

\begin{figure*}
\resizebox{\hsize}{!}{\includegraphics{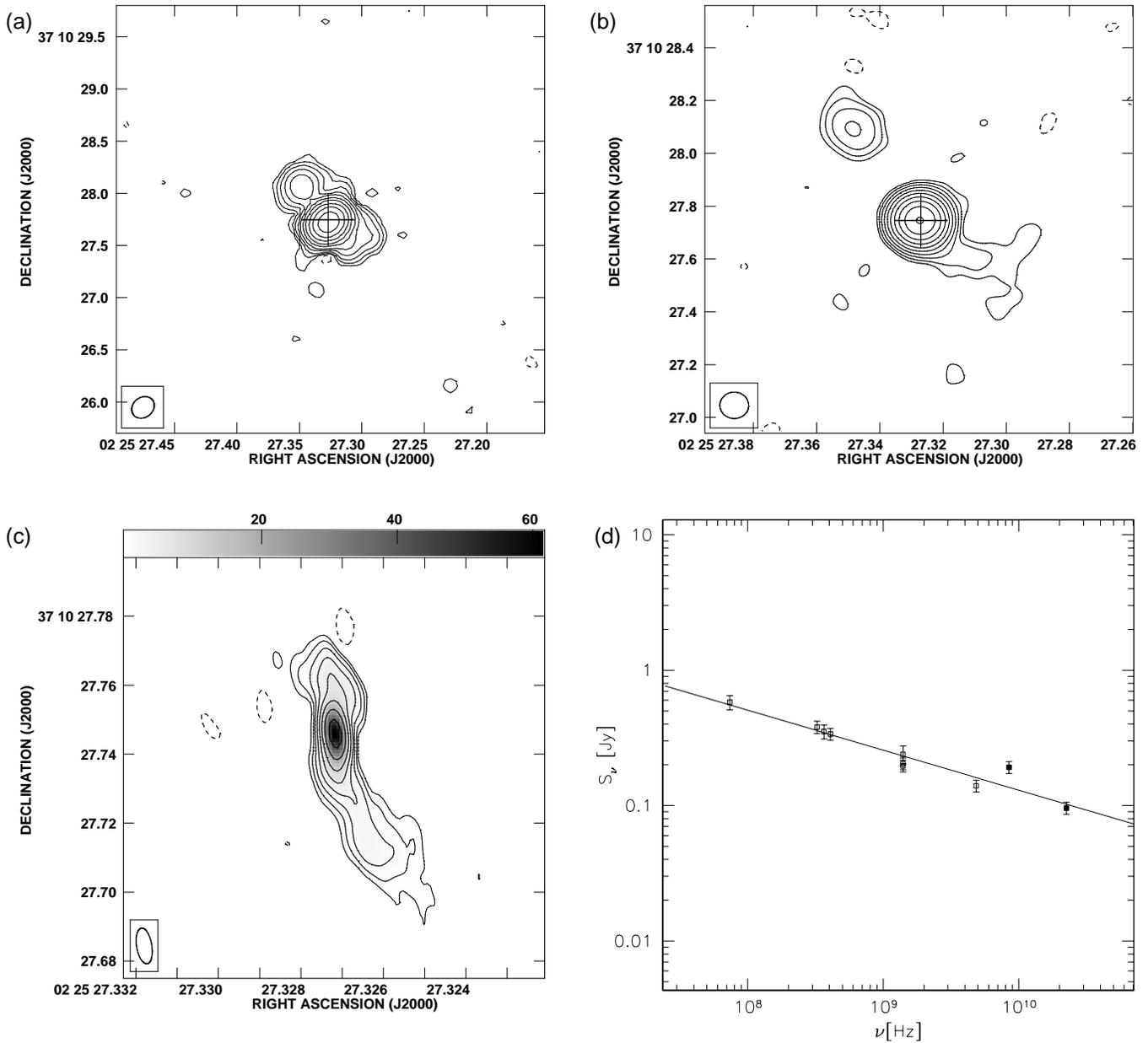}}
\caption{Images and spectrum for 0222+36. (a) VLA at 8.4 GHz, (b) VLA at 22.5 GHz, (c) VLBA at 1.6 GHz, (d) spectrum. Contours are traced at (1, 2, 4, ...) times the $3\sigma$ noise level; observational parameters (beam size, noise level, and peak brightness) are given in Table \ref{obs_para}. The cross in the VLA images indicates the position of the VLBA core. The integrated spectrum is derived from data in the present work and from the literature (see Sect.~\ref{results}). \label{fig0222+36}}
\end{figure*}

This source at $z=0.0334$ ($1\arcsec = 0.66$ kpc) has a flux density of 337 mJy
at 408 MHz, corresponding to a monochromatic total power of 10$^{23.9}$ W
Hz$^{-1}$.  Previous VLA observations at 1.4~GHz reveal a slightly resolved
morphology with a flat spectrum core surrounded by a halo extended over $\sim
8\arcsec$ \citep{fan86}. On parsec scales, EVN and low resolution VLBA
observations show an unresolved component with a flux density of $\sim$ 100 mJy
at 5 GHz \citep{gio01}.

Our 8.4 GHz VLA observations resolve the structure of 0222+36 into a core and
two components on either side (see Table \ref{res_vla}).  At 22 GHz the core is
the dominant structure and the two lobes are faint with an {\it S} shaped
structure (see Fig.~\ref{fig0222+36}).

The phase referenced VLBA image detects 102 mJy of flux density at 1.6 GHz in
total intensity.  No polarized flux is detected, which corresponds to a limit
on the fractional polarization of $<0.4\%$ at the core position.  The source is
two sided, with jets emerging in opposite directions along the north-south
axis. The unresolved central core has a flux density of 58.2 mJy.  Both jets
have a slightly bent path and become aligned with the kpc scale main axis at
$\sim 20$ mas from the core.  The resolved structure and the low core dominance
are in agreement with the moderately steep spectrum of the VLA core
($\alpha_{8.4}^{22} = 0.69 \pm 0.02$; see Tab.~\ref{res_vla}).

Both the parsec and kiloparsec scale morphology strongly suggests that the
source is oriented near the plane of the sky. At 10 mas from the peak, the
jet/counter-jet brightness ratio is $R=B_J/B_{CJ}=1.3$ ($B_J = 12.0\,$mJy and
$B_{CJ} = 9.5\,$mJy, with the main jet being the one pointing north); this
corresponds to $\beta \cos \theta \sim$ 0.05 which implies $\theta >
85^{\circ}$ if $\beta > 0.6$. Therefore, the source is not affected by
orientation effects and it has to be intrinsically small. If we consider the
largest angular extent of the source and deproject it with an angle of
$85^{\circ}$ we derive an intrinsic size of approximately 5.4 kpc.

\begin{figure}
\resizebox{\hsize}{!}{\includegraphics{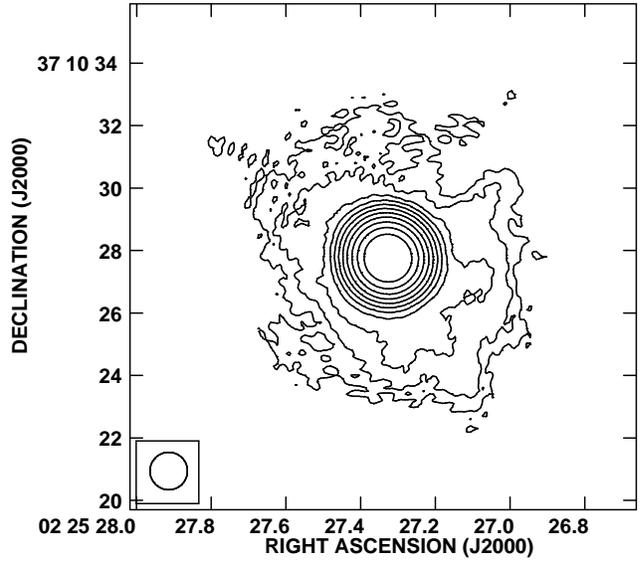}}
\caption{Image of 0222+36 at 5 GHz, VLA in ``A+B'' configuration. Contours are traced at (1, 2, 4, 8, ...) times the $3\sigma$ noise level. The lowest contour is $69\, \mu$Jy/beam, the peak is 209 mJy/beam. The beam is circular with a HPBW of 1.2\arcsec. \label{fig0222+36arch}}
\end{figure}

It is intriguing to try and understand the nature of the extended radio
emission surrounding the central components in the form of a $\sim 10$ kpc
halo. This halo is readily visible at low frequency \citep{fan86} but it is
completely resolved in our images.  For this reason, we reanalyzed VLA archive
data at 1.4~GHz (D array) and 5~GHz (in A + B configuration). The 1.4~GHz D
array data show that we are not missing a low brightness region more extended
than the halo, and the 5~GHz data allow us to derive the spectral index of the
halo region. We show in Fig.~\ref{fig0222+36arch} the 5~GHz A+B array image.

From a comparison of images at different resolution (see Fig.~\ref{fig0222+36}
and \ref{fig0222+36arch}), we do not see a clear connection between the inner
structure (VLBA and high frequency VLA) and the extended halo; on the other
hand, the VLBA structure is well connected to the source morphology as seen in
the 22 and 8 GHz images. We consider it unlikely that the halo is due to the
presence of extended lobes along the line of sight because this would imply a
large bend between the small and large scales with considerable fine tuning
required to produce such a symmetric, uniform, and circular extended emission.

\begin{figure*}
\resizebox{\hsize}{!}{\includegraphics{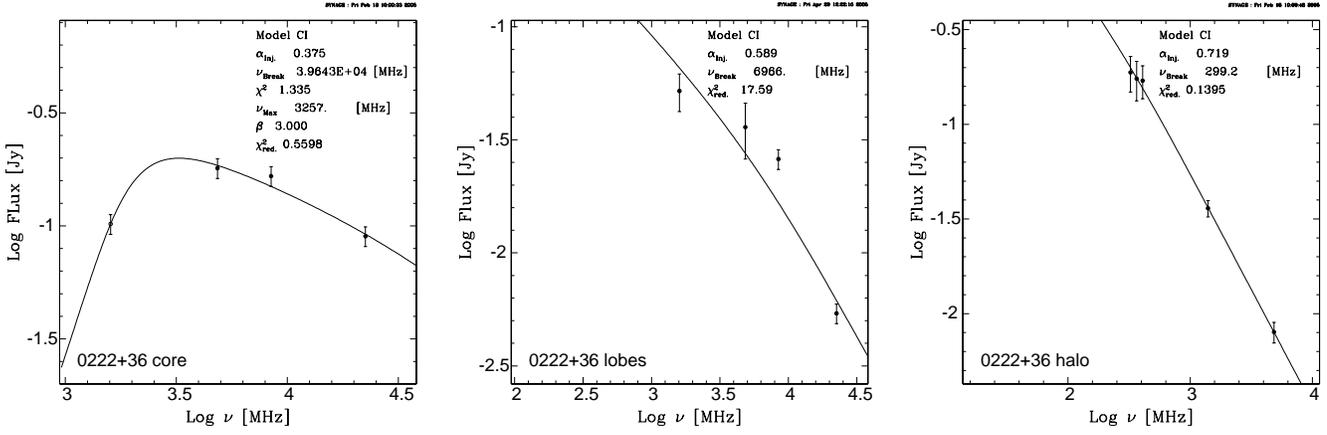}}
\caption{Spectra for the three components of 0222+36: (a) core, (b) lobes, (c) halo. the solid lines are fits with a model of continuous injection and low frequency self-absorption. The spectral points and the parameters resulting from the fits are presented in Table~\ref{tab0222+36sp}. \label{fig0222+36sp}}
\end{figure*}

We suggest that the halo could be due to the diffusion of relativistic
particles and magnetic field around the source, similar to the extended halos
seen in some spiral galaxies \citep{hum91}.  To better investigate this point,
we consider the radio spectrum of different components.  In Table
\ref{tab0222+36sp} we report flux density measurements between 74 MHz and 22.5
GHz; the data are also shown in Fig.~\ref{fig0222+36sp}, along with a best-fit
continuous injection model obtained using the Synage program \citep{mur96}. The
total spectrum (bottom right panel of Fig.~\ref{fig0222+36}) is complex but in
good agreement with the sum of spectra of different components
(Fig.~\ref{fig0222+36sp}). It is also worthwhile to remember that some
variability could be present and that the data at our disposal were not taken
simultaneously. However, we do not find any obvious inconsistency in the data,
and based on this fact we argue that the variability is insignificant.

We further point out that the core emission refers to the sub-arcsecond core
since we do not have multi-frequency VLBI data. For this reason, at 1.6 GHz we
used the total correlated flux in our VLBA data for the core flux density.
This value (102 mJy) is lower than the core flux density reported by
\citet{fan86}, which however does definitely include a contribution from the
lobes.  The resulting spectrum for the sub-arcsec core shows a turnover with a
maximum of 200 mJy at 3.3 GHz.  Assuming that the turnover is produced by
synchrotron self-absorption, we estimate, following \citet{mar87}, the average
magnetic field in the sub-arcsecond core to be in the range 2 -- 9 $\times$
10$^{-2}$ Gauss.

In our VLA images, both lobes show a steep spectrum between 8 and 22 GHz
($\alpha = 0.93 \pm 0.04$ and $0.97 \pm 0.04$ in the NE and SW lobe,
respectively). In order to obtain a better estimate of the physical parameters
in the lobes, we also consider lower frequency data. These data have been
obtained by subtracting the estimated flux density of the sub-arcsec core. For
this reason we give in Fig.~\ref{fig0222+36sp} and Table \ref{tab0222+36sp} the
spectrum of the two lobes together. This explains the larger dispersion and
uncertainty of this result. We estimate the average equipartition magnetic
field in the two lobes with standard assumptions: we consider a frequency range
from 10 MHz to 100 GHz, an equal amount of energy in heavy particles and in
electrons ($k = 1$), and that relativistic particles and magnetic fields occupy
the same volume ($\phi = 1$). Under these assumptions, we find
B$_{\mathrm{eq}}$ = 1.3 $\times$ 10 $^{-4}$ Gauss in the lobes and a break
frequency $\nu_{\mathrm{br}} \sim 9$ GHz.

On larger scales, we derive a flux density of the halo from the archival data
in A and B configuration at 5 GHz (see the image obtained combining A+B data in
Fig.~\ref{fig0222+36arch}). Similarly to the case of the lobes, we derive the
equipartition magnetic field in the extended halo. The spectrum of this region
is straight and steep ($\alpha \sim$ 1.2), with a low break frequency
$\nu_{\mathrm{br}} \la 0.3$ GHz. In this region we have B$_{\mathrm{eq}} \sim
6.9 \times 10^{-6}$ Gauss.

From the spectral index information and the equipartition magnetic field, we
estimate the radiative age in the lobes and in the halo region. We find that
the lobes are 4.5 $\times$ 10$^5$ yrs old and the extended halo is $\ga 1.3
\times 10^8$ yrs old. Therefore, we have in this source a young structure
(inner lobes) surrounded by an older symmetric and diffuse region. Since there
is evidence of a change of the jet direction in the inner young region, we can
speculate that the jet orientation is rotating because of instabilities of the
AGN.  This could explain the small size of the source because an unstable jet
did not allow the growth of a large scale radio galaxy.  In this case the old
round halo could be due to the diffusion of radio emission during the orbit of
the inner structure. From VLBA and high resolution VLA images we can estimate
that the jet rotated by $\sim$ 40$^\circ$ in 5 $\times$ 10$^5$ yrs so that in
10$^8$ yrs (the halo age) it could have done many complete orbits. Higher
sensitivity A+B+C data at 8.4 GHz might be able to follow the radio structure
from the lobes to the halo to test this hypothesis.

We can also compare our estimate of the spectral age of the lobes to a
reasonable dynamic estimate based on their size. Assuming that the two lobes
separation velocity is $\sim$ 0.1 c, we derive a dynamic age $\sim 2 \times
10^4$ yrs. This age is 10 times lower than the radiative age, but such
differences have been found in other CSOs \citep[e.g. 4C31.04,][]{gir03} and
could be a result of the many assumptions going into both the spectral and
kinematic age estimates (e.g., constant advance velocity and equipartition
conditions).

\subsection{0258+35 (NGC 1167)} 

\begin{figure*}
\resizebox{\hsize}{!}{\includegraphics{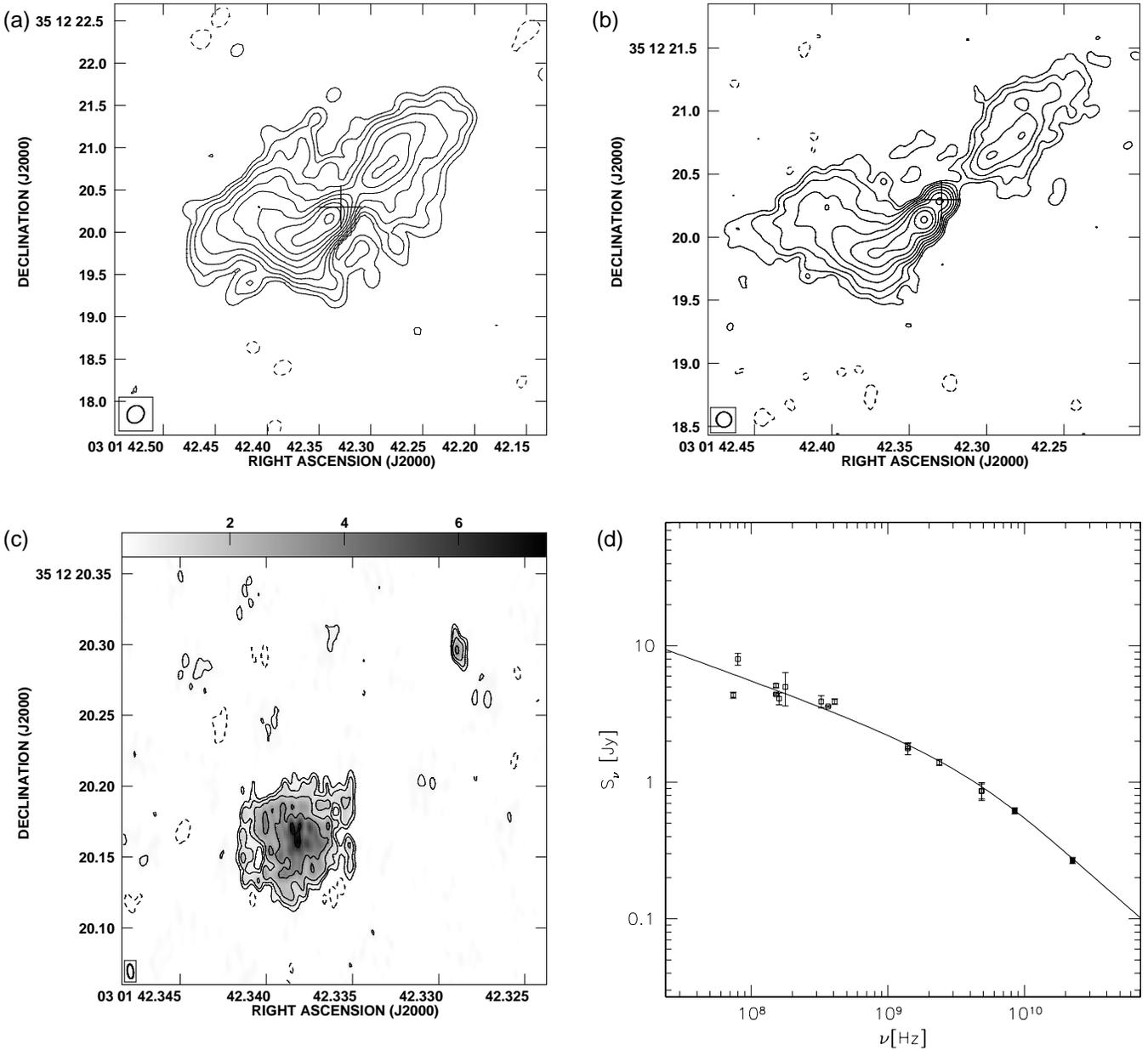}}
\caption{Images and spectrum for 0258+35. (a) VLA at 8.4 GHz, (b) VLA at 22.5 GHz, (c) VLBA at 1.6 GHz, (d) spectrum. Contours are traced at (1, 2, 4, ...) times the $3\sigma$ noise level; observational parameters (beam size, noise level, and peak brightness) are given in Table \ref{obs_para}. The cross in the VLA images indicates the position of the VLBA core. The integrated spectrum is derived from data in the present work and from the literature (see Sect.~\ref{results}). \label{fig0258+35}}
\end{figure*}

At $z = 0.0165$ ($1\arcsec = 0.34$ kpc), this source has a radio power at 408
MHz of $10^ {24.37}$ W Hz$^{-1}$. It is optically classified as a Seyfert 2
galaxy \citep{ho97}, and in the radio it has been previously studied and
classified as a compact steep spectrum (CSS) source by \citet{san95}, even
though the total spectrum is only moderately steep: $\alpha_{0.08}^{22} =
0.54$. In contrast with most CSS sources, 0258+35 shows plume-like lobes
without prominent hot-spots.  Previous studies include the global VLBI image at
5 GHz in \citet{gio01}, which reveals a compact component with a short jet-like
structure and an excess of emission on the short spacings.

Our VLA images are shown in Fig.~\ref{fig0258+35} (top panels). The 8.4 GHz
image shows a morphology in good agreement with previous images. Thanks to the
high sensitivity of the VLA at this frequency, an extended diffuse emission
surrounding the inner radio structure is detected.  In the 22 GHz image, thanks
to the better angular resolution, the source peak is resolved into two
components, one of which is fainter and more central, while the other is
stronger and located at the beginning of the SE lobe.

Our phase-referenced VLBI data show a faint compact component identified with
the nuclear source and an extended ``blob'' at $\sim 0.1\arcsec$. As shown by
the cross overlaid on the VLA images, the compact parsec scale core is located
in the central component at the base of the south-east jet-like feature. This
component is easily identifiable in the 22 GHz image, while in the 8 GHz map
the core is confused with a bright jet component. Besides the faint compact
core ($S_{\mathrm{c}} = 7.6$ mJy), the largest fraction of flux density (240
mJy) in the VLBI image at 1.6 GHz is contained in a bright diffuse region
coincident with the peak of the VLA images.  The flux density and size of the
components are given in Table \ref{res_vlba}.  We wonder about the nature of
this region because of its morphology and the lack of a connection with the
parsec-scale core.  It could be the result of a burst of activity of the
central AGN, in which case some moderate Doppler boosting is necessary to
explain the absence of a similar counterjet on the opposite side.  From the jet
to counter-jet ratio in the 22 GHz VLA map measured at $\sim 0.5\arcsec$ from
the core, we find that the source has to be oriented at an angle $<$
70$^\circ$; if the source has subarcsecond relativistic jets, the orientation
is in the range $60^\circ \la \theta \la 70^\circ$, in agreement with the
optical classification of the parent galaxy as a Seyfert 2. From the VLBI image
and using the knot brightness we find $R=B_J/B_{CJ} \ge 7.5/0.13 \sim 58$,
resulting in $\beta\cos\theta>0.67$, and a viewing angle $40^\circ \la \theta
\la 50^\circ$, if $\beta \ge 0.9$. However, we cannot rule out the possibility
that the knot was produced in a peculiar episode and the jet/counterjet ratio
is not meaningful.

The 22 GHz VLA image has about the same resolution ($\sim 80$ mas) as the
combined EVN+MERLIN 1.6 GHz map shown in \citet{san95}, and we can estimate the
spectral index for the core ($\alpha \sim 0.0$) and main jet component ($\alpha
\sim 0.4$).  A flat core spectral index is confirmed by our VLA images, albeit
a slight misalignment between the two images prevents an unambiguous core
identification only from our data.  The spectrum is still flat in the central
region and gradually steepens to become almost constant (0.6) in the inner
bright jet-like structure in both lobes.  The surrounding diffuse emission is
steep: 1.0 -- 1.5.  The integrated spectral index at high frequency is
$\alpha_{8.4}^{22} = 0.88 \pm 0.01$.

Since the knot is well resolved in our VLBA image we derive its opening
angle. We measure an angle of 26$^\circ$, which is close to the intrinsic
opening angle if the source is oriented near the plane of the sky. A free
expanding jet is expected to show an intrinsic opening angle of $\sim 1/\gamma$
\citep[see, e.g.][]{sal98}; under this assumption, we estimate a Lorentz factor
$\gamma \sim$ 2.4 which corresponds to $\beta \sim$ 0.9.

\begin{figure*}
\resizebox{\hsize}{!}{\includegraphics{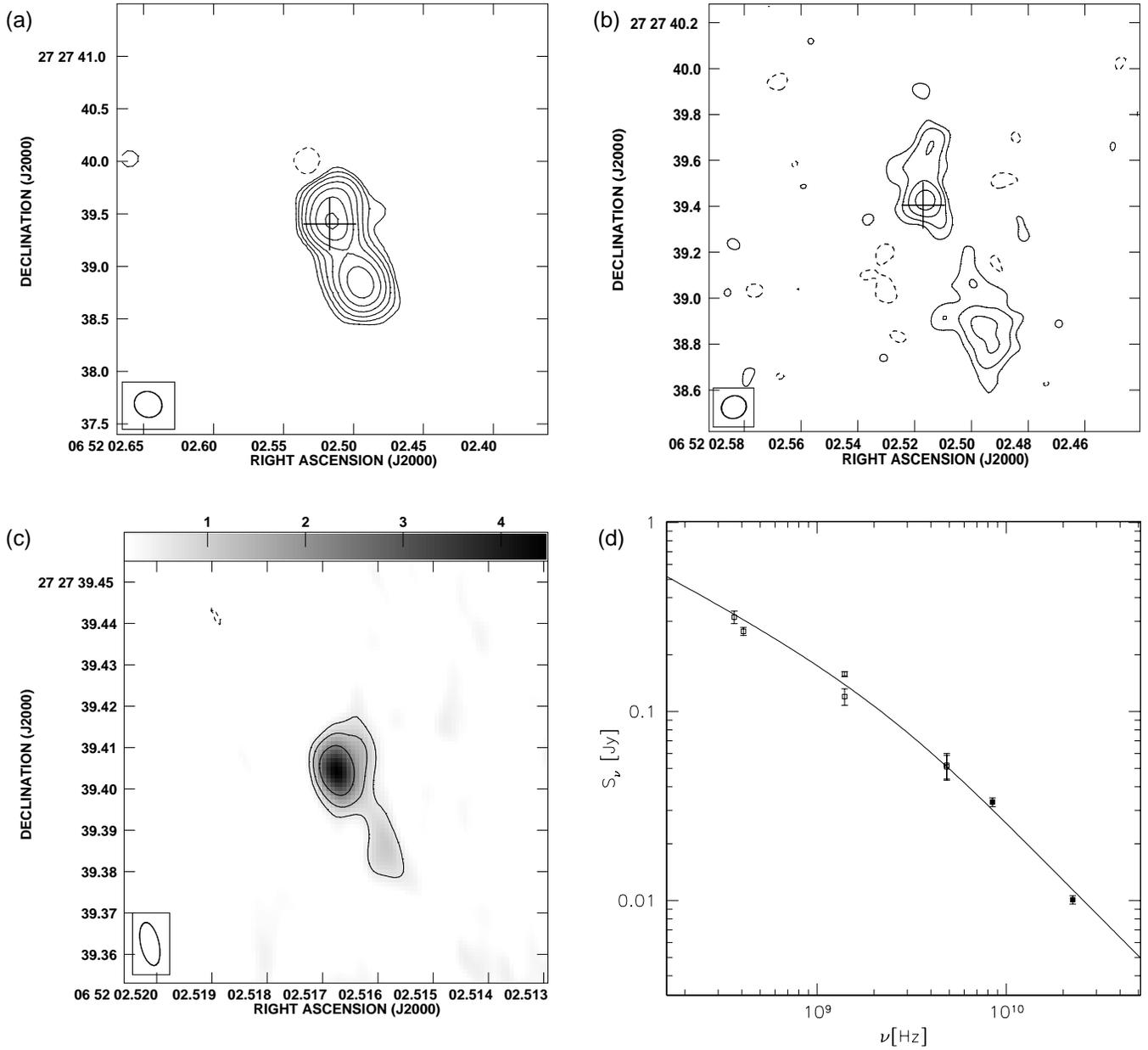}}
\caption{Images and spectrum for 0648+27. (a) VLA at 8.4 GHz, (b) VLA at 22.5 GHz, (c) VLBA at 1.6 GHz, (d) spectrum. Contours are traced at (1, 2, 4, ...) times the $3\sigma$ noise level; observational parameters (beam size, noise level, and peak brightness) are given in Table \ref{obs_para}. The cross in the VLA images indicates the position of the VLBA core. The integrated spectrum is derived from data in the present work and from the literature (see Sect.~\ref{results}). \label{fig0648+27}}
\end{figure*}

The source structure suggests that in the data at 5 GHz \citep{gio01} the main
peak is produced by the nuclear source with the beginning of the main jet also
visible.  In this case the VLBI core has an inverted spectrum, free-free or
self-absorbed (26 mJy at 5 GHz and 7.5 mJy at 1.6 GHz).

We have estimated the average equipartition magnetic fields in the source,
assuming an uniform brightness in the source volume. This can be considered a
good approximation, since the compact structures (core and knot) are only $\sim
20\%$ of the total flux density at 5 GHz.  We estimate B$_{eq}$ $\sim$ 9
$\times$ 10$^{-5}$ Gauss.  With this estimate and using the break frequency
found in the total spectral index distribution (4.6 GHz) we estimate an age of
9 $\times$ 10$^5$ yrs for this source. Of course, this is an average age and we
expect that external diffuse regions are older with respect to the innermost
region.

In light of these results, we speculate that this source might not grow to
become a kiloparsec scale radio galaxy. No final hot spots demarcating the ends
of the jets are visible and the source structure appears to strongly interact
with the ISM as shown by the large bending of the arcsecond structure of the SE
lobe and the presence of a surrounding low brightness extended structure in the
VLA images. We note that in this source the inner jet direction appears to be
constant, moreover no amount of bending is visible in the NW lobe, therefore
the source structure on the large scale should be related not to the inner BH
motion but to interaction with the ISM. We note that the estimated radiative
age should imply a larger source size even allowing for a low lobe advance
velocity.

\subsection{0648+27} 

\begin{figure*}
\resizebox{\hsize}{!}{\includegraphics{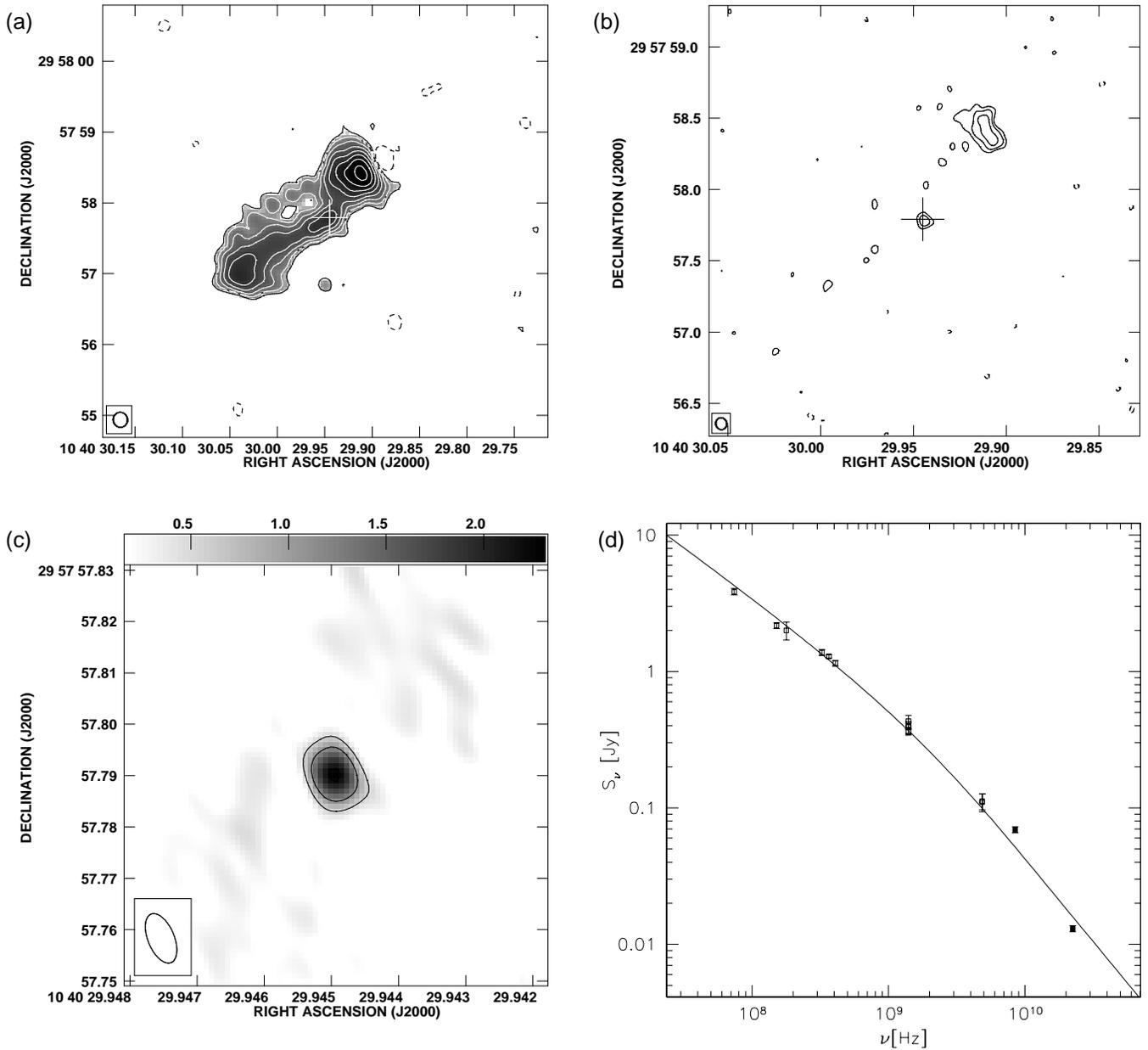}}
\caption{Images and spectrum for 1037+30. (a) VLA at 8.4 GHz (grey scale between 0.1 and 10 mJy/beam), (b) VLA at 22.5 GHz, (c) VLBA at 1.6 GHz, (d) spectrum. Contours are traced at (1, 2, 4, ...) times the $3\sigma$ noise level; observational parameters (beam size, noise level, and peak brightness) are given in Table \ref{obs_para}. The cross in the VLA images indicates the position of the VLBA core. The integrated spectrum is derived from data in the present work and from the literature (see Sect.~\ref{results}). \label{fig1037+30}}
\end{figure*}

This object ($z = 0.0414$, corresponding to 0.82 kpc/\arcsec,
$P_{408\,\mathrm{MHz}} = 10^{24.02}$ W Hz$^{-1}$) is only slightly extended at
the lowest frequencies \citep{par86}. It was resolved into a double source
extended about 1\arcsec\, with VLA observations at 8.4 GHz by \citet{mor03};
they also detect a large amount of H{\sc i} using the Westerbork Synthesis
Radio Telescope (WSRT, $M_{\mbox{H{\sc i}}} = 1.1 \times 10^{10} M_\odot$).
However, \citet{mor03} are not able to identify a core and consequently
interpret the radio structure in terms of a pair of symmetric lobes.

Our 8.4 GHz VLA data (Fig.~\ref{fig0648+27}, top left panel) confirm the
structure as a double, with flux densities of 19 and 14 mJy in the northern and
southern components, respectively, in agreement with the data from
\citet{mor03}. However, both components are resolved at 22 GHz (see top right
panel in Fig.~\ref{fig0648+27}), and a compact feature emerges in the north
with a flux density of 2.5 mJy. Its small size ($<0.07\arcsec$) and flatter
spectral index suggest that this component is actually the core, with emission
on either side. See Table~\ref{res_vla} for a list of our data and spectral
index measurements for the two lobes.

Our phase referenced VLBI data lend strong support to this scenario.  In fact,
the emission in the VLBA image is located in the vicinity of the VLA peak in
the northern lobe. The total flux density is only 12.8 mJy, with a peak of 4.4
mJy/beam. A faint jet-like structure is visible to the south-east, although the
signal-to-noise is very poor and it could be spurious. In any case the
difference between the total correlated flux and the peak flux density in the
VLA image suggests the presence of some extended emission on intermediate
scales.  We note that the spectral index between the 22 GHz VLA unresolved
component and the total VLBA correlated flux is $\alpha^{22}_{1.6} = 0.47 \pm
0.03$ (see Table~\ref{res_vlba}). It is possible that this component is a jet
knot and the core is at the extreme northern end. The core might be free-free
or self-absorbed at 1.6 GHz, so that the VLBA does not pick out the center of
activity.

The source flux density is dominated by the extended emission. The total
spectrum of the source between 325 MHz and 22 GHz has an index $\alpha \sim
0.8$, with a hint of steepening at high frequency ($\alpha_{8.4}^{22} = 1.22
\pm 0.02$). A measurement at 5 GHz \citep[$S_\mathrm{5\,GHz} = 213$
mJy,][]{ant85} deviates significantly from the other measurements and was not
included in the spectrum. If it is not due to an error, then it is difficult to
understand this value.

The two lobes do not show any evidence for jet-like structure or the presence
of hot spots.  The lobe structure is relaxed with a relatively steep spectrum.
The South lobe is at a greater distance with respect to the core, but we
ascribe this asymmetry to a difference in the ISM and not to a relativistic
effect. We note also that the lack of prominent jet structures in the parsec
scale image suggests that no relativistic jet structure is present or that the
source is on the plane of the sky. The faintness of the nuclear emission also
argues against a Doppler boosted jet.

We have estimated the average equipartition magnetic field in this source to be
H$_\mathrm{eq} \sim 95 \times 10^{-6}$ Gauss.  Using the break frequency (2.9
GHz) estimated from total flux density measures we can estimate a minimum age
for this source of about 1 Myr ($9.9 \times 10^5$ years).  We expect that the
external lobe regions are much older, confirming that this source is confined,
and is expected to remain compact similar to NGC 4278 \citep{gir05} despite its
relatively higher total radio power and larger size.  The connection between
the small size of the radio emission and the presence of a major merger in this
galaxy about 10$^9$ years ago, with the presence of a large amount of H{\sc i}
in this galaxy \citep{mor03}, is remarkable.

\subsection{1037+30}

\begin{figure}
\resizebox{\hsize}{!}{\includegraphics{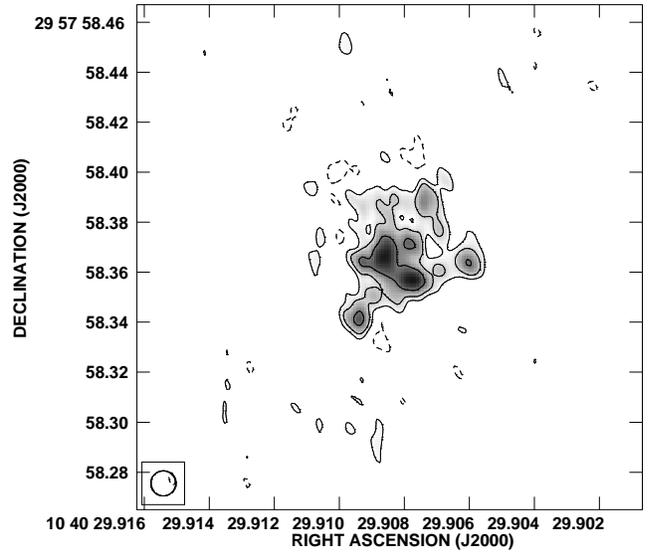}}
\caption{Phase referenced VLBA image of the hot spot region of 1037+30. Contours are traced at (1, 2, 4, 8, ...) times the $3\sigma$ noise level. The lowest contour is 0.3 mJy/beam and the peak is 5.6 mJy/beam; the gray scale range is between 0.9 and 6 mJy/beam. The beam is circular with a HPBW of 10 mas. \label{fig1037+30hs}}
\end{figure}

\begin{figure*}
\resizebox{\hsize}{!}{\includegraphics{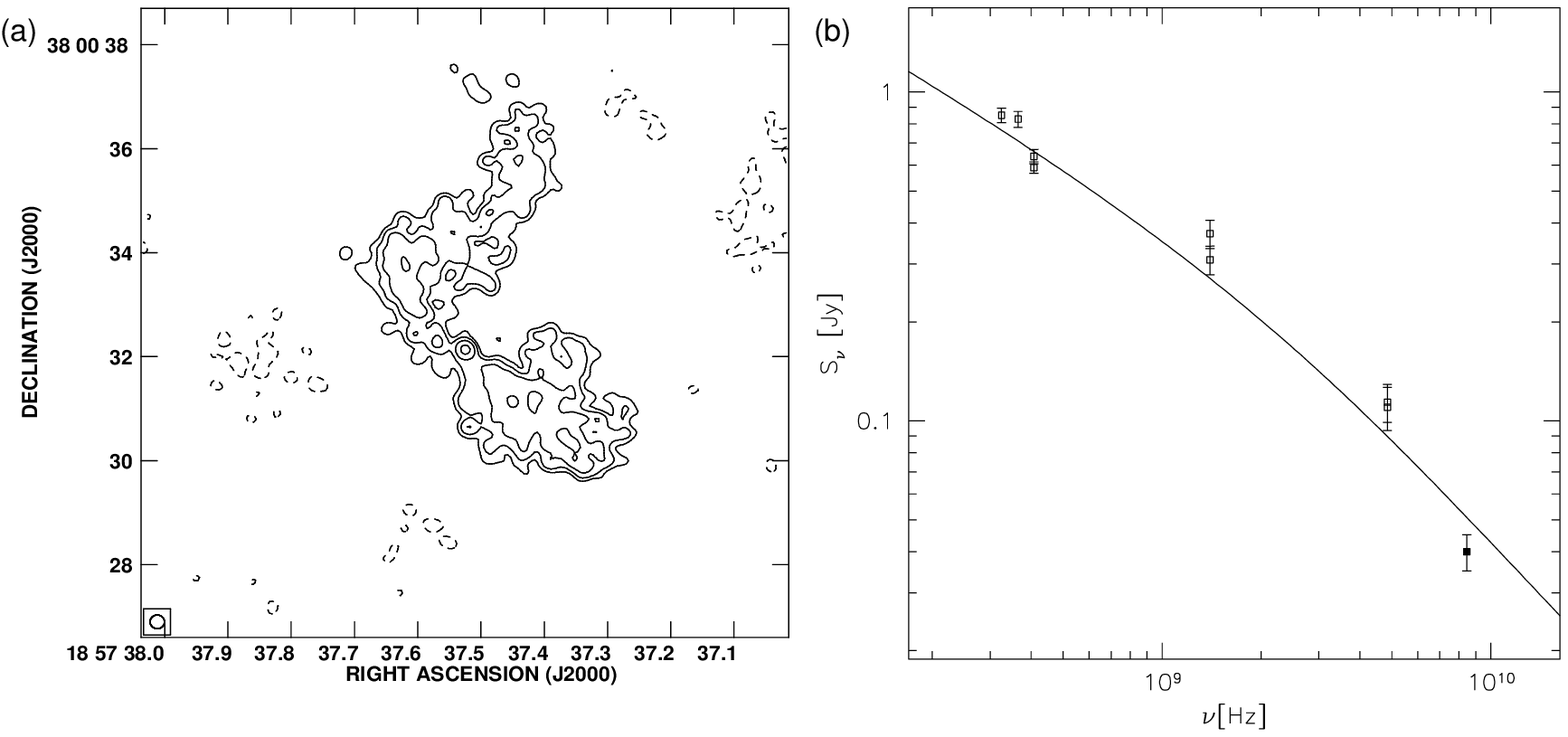}}
\caption{The VLA image and spectrum for 1855+37. (a) VLA at 8.4 GHz, (b) spectrum. Contours are traced at (1, 2, 4, ...) times the $3\sigma$ noise level; observational parameters (beam size, noise level, and peak brightness) are given in Table \ref{obs_para}. The source is not detected with the VLA at 22 GHz nor at 1.6 GHz with the VLBA. The integrated spectrum is derived from data in the present work and from the literature (see Sect.~\ref{results}). \label{fig1855+37}}
\end{figure*}

Located at $z =0.0911$ ($1 \arcsec = 1.70$ kpc), 1037+30 has
$P_{\mathrm{408\,MHz}} = 10^{25.37}$ W Hz$^{-1}$. It is only slightly resolved
at 1.4 GHz \citep{fan86} and not detected in any VLBI observations to date
\citep{gio05}. In the optical, 1037+30 is identified with the brightest galaxy
in the cluster Abell 923.

Our VLA images are shown in the top panels of Fig.~\ref{fig1037+30}. The 8.4
GHz image reveals an edge-brightened structure, with complex sub-structures:
jets, and lobes with hot spots. In the 22 GHz image only a point-like
component, probably the core, and the resolved NW hot spot are evident.  As in
other objects, the accurate phase-referenced VLBI image provides confirmation
for the identification of the core which appears as a faint central component
in the 8.4 GHz image.

The south-eastern jet is clearly visible in the 8.4 GHz VLA image, and is both
longer and better defined than the jet in the opposite direction. While the jet
brightness and length suggest that the SE is the approaching side, the NW hot
spot is much brighter (with a peak of 13 vs 2.3 mJy/beam at 8.4 GHz). It is
possible that the ISM density is irregular and that the brightness asymmetry is
due to interactions with an inhomogeneous ISM.

The NW hot spot has a peculiar morphology, with a very sharp edge.  We
interpret this structure in terms of a back-flow due to a strong interaction of
the NW jet with the ISM. This could also explain the compressed appearance of
the NW hot spot, its higher brightness, and the ``bridge'' connecting the NW
hot spot with the SE hot spot.  Since the jet to counterjet brightness ratio in
the 8.4 GHz image is not high, we expect that the source is oriented at a large
angle with respect to the line of sight.  This orientation is in agreement with
the non-detection of the jet at 22 GHz and at 1.6 GHz by the VLBA.  This
orientation is also consistent with the low core dominance, since relativistic
jets at a large angles will have a Doppler factor $<<$ 1.

The VLBA data show a clear detection of a 4 mJy core (see
Table~\ref{res_vlba}). Faint, diffuse emission is detected on the shortest
baselines, although it is not possible to image it properly. At the very limit
of the VLBA field-of-view, we detect radio emission from the north west hot
spot region. Our image (Fig.\ref{fig1037+30hs}) shows a resolved structure
$\sim 60$ mas in size in agreement with the 22 GHz image. The total flux
density at 1.6 GHz in this region is $\sim 40$ mJy. Deeper images are necessary
to see the connection to the jet.

The size and morphology of this source are in agreement with the definition of
CSO sources. The jet interaction with the ISM is well defined. We estimated a
dynamical age assuming a lobe expansion velocity of 0.2c and find an age of
$4.5 \times 10^4$ yrs assuming for the size the distance from the core of the
SE hot spot. We do not estimate for this source a synchrotron age: the complex
radio structure makes any estimate of an average H$_{eq}$ unrealistic, and the
contribution of different components to the total spectrum is difficult to
weigh. The overall break frequency is however around 2.7 GHz.

We note that because of its total radio power we expect that 1037+30 should
evolve into an extended FR I radio galaxy. This source, like 0116+31
\citep[4C\,31.04,][]{gir03}, is also one of a number of low power radio sources
with a clear CSO morphology.

\subsection{1855+37}

Located at a redshift $z=0.0552$ ($1\arcsec = 1.07$ kpc), 1855+37 has a total
power $P_\mathrm{408\,MHz} = 10^{24.65}$ W Hz$^{-1}$. Compact in the NVSS, this
object is resolved into a triple source with higher resolution VLA observations
at 1.4 GHz \citep{fan86} and 5 GHz \citep{mor97}, where it also shows
polarization. We identify 1855+37 with the brightest member of the galaxy
cluster CIZA J1857.6+3800, discovered on the basis of X-ray data by
\citet{ebe02} in the zone of avoidance. A discussion of the X-ray properties of
1855+37 on the basis of ROSAT PSPC data is presented in \citet{can99} and
\citet{wor00}.

Our observations at 8.4 GHz (Fig.~\ref{fig1855+37}) resolve the flux density of
$\sim 40$ mJy into a weak, diffuse emission extended over $\sim 7\arcsec$.  The
structure is two sided, with a weak central component ($S_\mathrm{core} = 0.57$
mJy) and large, faint lobes. The source is not detected at 22 GHz; no
significant peak is found in this region in the VLBA image.

Given the low power of the core, the lack of visible jets, and the rather steep
spectral index, it is possible that the activity in this source is fading
away. Note that in all previous observations, typically at lower frequency
\citep{fan86,mor97}, the source was largely dominated by the extended
lobes. For this reason, it was not possible to pinpoint the location of the
nuclear activity, whose extreme weakness has therefore gone unnoticed. Our 8.4
GHz observations reveal for the first time the exact position and flux density
of the core, which is remarkably weak ($S_\mathrm{core}/S_\mathrm{tot} =
0.017$). This also explains the lack of detection of nuclear activity in 22 GHz
VLA data and in our VLBI observations.  We note also that according to the
correlation between the core and total radio power \citep[see e.g.][]{gio01}
the nuclear power in this source is too low even assuming a large jet velocity
and a huge Doppler de-boosting.  We therefore propose this galaxy hosts a dying
radio source.  If the nucleus of this galaxy is ceasing its activity we would
not expect that this source would significantly grow any larger than it is now
($\sim 7$ kpc).

We note that the present radio structure is very similar to that found in NGC
4874 (the brightest galaxy in the Coma cluster) by \citet{fer85}. However, in
NGC 4874 the presence of a core and of bright regions in the two lobes
suggests ongoing nuclear activity.

The few data points available in the literature, combined with our new
measurements at 8.4 GHz and limit at 22.5 GHz, yield a break frequency
$\nu_\mathrm{br} \sim 2.5$ GHz. The equipartition magnetic field is $\sim$ 2.1
$\times$ 10$^{-5}$ Gauss and the corresponding radiative source age is 9.7
Myrs. Since this source is inside an X-ray cluster we expect that its small
size with respect to the radiative age can also be due to confinement from the
IGM gas \citep[see also][]{wor00}.  Also the source shape, similar to wide
angle tail radio sources, could be due to relative motions between thermal gas
and the source.

\section{Discussion}

\begin{figure}
\resizebox{\hsize}{!}{\includegraphics{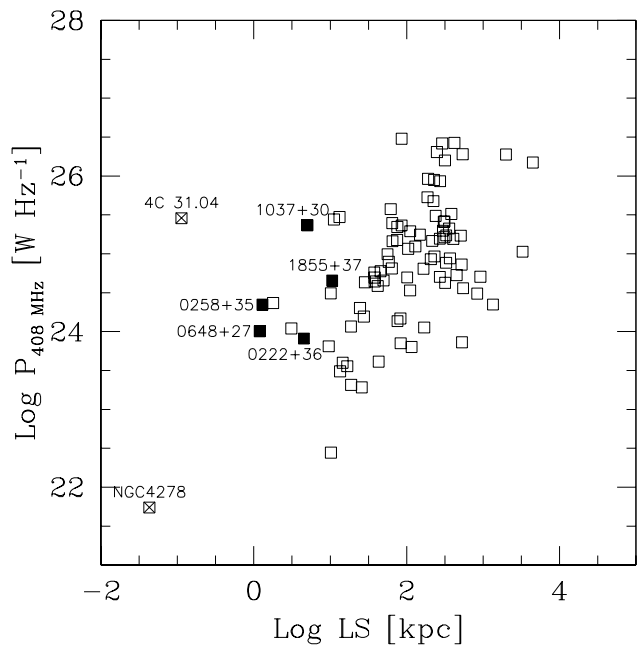}}
\caption{Radio power vs linear size diagram for sources in the Bologna Complete Sample \citep[BCS,][]{gio05}. The filled squares indicate sources studied in the present work, while the starred symbols are for the two other compact sources in the sample: 0116+31 \citep[4C 31.04,][]{gir03} and 1217+29 \citep[NGC 4278,][]{gir05} \label{ls}}
\end{figure}

We present new VLA and VLBA results for 5 Low Power Compact Radio Sources. Our
high resolution images resolve all sources. The combined information of
flat/inverted spectral indices ($-0.2 \la \alpha_{1.6}^{22} \la 0.5$) and
compactness unambiguously identify radio cores in four of them. In one case
(1855+37), we have a compact component at 8.4 GHz, which is a reasonable core
candidate; however, a detection at another frequency would be desirable to
confirm the classification. Most sources are symmetric, at least on arcsecond
scales; 0222+36 is also clearly symmetric on milliarcsecond scales. In general,
the objects do not show any evidence for strong beaming effects (e.g. strong,
one-sided jets, dominant cores). Thus, these sources are most likely close to
the plane of the sky. We determine accurate values for their (projected) linear
size, which must be, therefore, within a factor of a few from their intrinsic
deprojected dimensions.

We show in Fig.~\ref{ls} a radio power vs linear size diagram for the 95
sources in the Bologna Complete Sample (BCS). Radio power data are taken from
the B2 survey at 408 MHz via the on-line VizieR service\footnote{{\tt
http://vizier.u-strasbg.fr/viz-bin/VizieR}} \citep{och00} and linear sizes from
several works in the literature\footnote{The largest part of the data is drawn
from \citet{fan87} and from ``An Atlas of DRAGNs'' for sources that are part of
the 3CRR sample (see {\tt http://www.jb.man.ac.uk/atlas/}); a few individual
sources are better studied in other references: 0116+31
\citep[4C\,31.04,][]{gir03}, 0836+29B \citep{fan86}, 1144+35 \citep{gio99},
1217+29 \citep[NGC\,4278,][]{gir05}, 1257+28 \citep[NGC\,4874,][]{fer85},
1557+26 \citep[from the FIRST survey,][]{bec95}, 1652+39A \citep[Mkn
501,][]{cas99}}. The solid points denote the five LPC sources in the subsample
considered in the present work with accurate linear size measurements. We
confirm that this class of radio sources is characterized by a linear size $<$
10 kpc.  The starred points indicate the two smallest sources in the BCS
($LS<0.1$ kpc), one of which is a genuinely young CSO \citep[4C
31.04,][]{gir03}, while the other is associated with a less active LINER galaxy
\citep[NGC 4278,][]{gir05}. From the distribution shown in the diagram of
Fig.~\ref{ls}, we note that the small size of LPCs is not related to the low
power of the radio source. In fact, there is also a large number of extended
($LS > 10^2$ kpc) radio galaxies with a similarly low radio power.

Our new LPC sources are more extended than classical young radio sources like
the CSOs \citep{gug05}, and could represent the intermediate age sources in
between CSOs and classic kpc scale radio galaxies.  The youth scenario has the
problem of the lack of objects filling in the gap between sources with age
$<10^4$ yrs and $>10^7$ yrs; in particular, there seem to be too many bright
young sources. However, this problem can be solved if the radio sources dim in
luminosity as they grow up; for example, assuming equipartition conditions and
expansion losses, \citet{beg96} reconciles the observational data with a model
that predicts that the radio luminosity should decrease with size, roughly as
$P_R \propto LS^{-1/2}$.  According to the evolution scenario young radio
sources are then expected to be small and quite luminous.

In order to better understand radio source evolution and the radio power vs
linear size diagram, we need to know the properties of compact radio
sources. It is also important to point out that the constraints posed by the
number counts per linear size range may be misleading. In fact, a number of
explanations have been invoked, for example contamination from core-jet
Doppler-boosted sources \citep{tin05} and a significant fraction of either
frustrated or short lived sources \citep{rea94,gug05}.

To investigate these possibilities we are studying in detail compact (LS $<10$
kpc) sources in our BCS, which are typically low power ones. If we consider the
data presented here for 5 sources and include also the two other sources
previously studied \citep[4C\,31.04, NGC 4278;][]{gir03,gir05}, we are
presented with a variety of behaviors. As a matter of fact, the individual
study of each source suggests that all cases are possible and we have evidence
in a number of cases that sources are not able to grow because of an
underpowered core or a jet instability, or because they are dying.  On the
other hand, we still have only a few cases in which there is clear evidence of
a source which is growing, with significant interaction between the jet and the
ISM. In particular, among sources studied here, we have:

-- One source with evidence of a growing jet structure interacting with the ISM
(1037+30) with an estimated kinematic age of 5 $\times$ 10$^4$ yrs.

-- A peculiar source in which an old structure and a younger one coexist
(0222+36, in which the estimated radiative age of the halo is $\sim 10^8$ yrs,
while the lobes are $10^4 - 10^5$ yrs old).  We interpret this structure as due
to a jet instability which does not allow it to create large scale lobes. The
old halo is permeated by relativistic electrons which slowly escape from the
younger region (lobes) and diffuse into the surrounding volume.

-- Three sources that are small because they are short lived or frustrated. In
0258+35, which has no hot spots and resembles a FR I radio galaxy, the VLA
structure does not show any evidence of interaction between expanding jets and
the ISM.  The VLBA structure (a faint core and an isolated more powerful knot)
suggests variable levels of activity. In 0648+27 and 1855+37 we find a low
power nuclear region and little or weak extended emission, suggesting that they
are possibly frustrated and/or dying. In the H{\sc i} rich 0648+27, we detected
a nuclear source with the VLBA and the estimated radiative age is $\sim$ 10$^5$
yrs, while in 1855+37 the nuclear source is no longer active and the source
radiative age is 10$^7$ yrs. In general, the dynamic ages estimated from their
large-scale structure and assuming typical lobe advance velocities are about
one order of magnitude lower than the spectral estimates. Unless there is some
misleading assumption (e.g., equipartition conditions do not apply), the
advance velocities ($<< 0.1c$) have to be significantly lower than those
measured in CSOs.

\section{Conclusions}

Despite the poor statistics, our preliminary studies indicate that multiple
causes can produce sources in the LPC class.  In addition to flat or inverted
spectrum sources dominated by projection effects as BL-Lacs, a small size can
stem from: youth (4C 31.04 and 1037+30), instabilities in the jets (in space,
as in 0222+36, or time, as in 0258+35), frustration (0648+27), a premature end
of nuclear activity (1855+37), or just a very low power core (NGC 4278). A more
detailed discussion will appear in a future paper about the BCS. The study of
the BCS, as a well defined complete sample, will allow us to make a statistical
study to derive the source evolution.  From the number of sources in different
evolutionary stages it should be possible to estimate the duration and
probability of different stages in radio source life.

\begin{acknowledgements}
We thank the referee Dr. C. Stanghellini for a prompt and useful report.  The
National Radio Astronomy Observatory is a facility of the National Science
Foundation operated under cooperative agreement by Associated Universities,
Inc. This research has made use of the NASA/IPAC Extragalactic Database (NED)
which is operated by the Jet Propulsion Laboratory, Caltech, under contract
with NASA and of NASA's Astrophysics Data System (ADS) Bibliographic
Services. This material is based upon work supported by the Italian Ministry
for University and Research (MIUR) under grant COFIN 2003-02-7534.
\end{acknowledgements}

\bibliographystyle{aa}

\begin{thebibliography}{}

\bibitem[Antonucci(1985)]{ant85} Antonucci, R.~R.~J.\ 1985, \apjs, 59, 499
\bibitem[Becker et al.(1995)]{bec95} Becker, R.~H., White, R.~L., \& Helfand,
D.\ 1995, \apj, 450, 559
\bibitem[Begelman(1996)]{beg96} Begelman, M.~C.\ 1996, Cygnus A -- Study of a
Radio Galaxy, 209
\bibitem[Blundell et al.(1999)]{blu99} Blundell, K.~M., Rawlings, S., \&
Willott, C.~J.\ 1999, \aj, 117, 677
\bibitem[Canosa et al.(1999)]{can99} Canosa, C.~M., Worrall, D.~M., Hardcastle,
M.~J., \& Birkinshaw, M.\ 1999, \mnras, 310, 30
\bibitem[Capetti et al.(2002)]{cap02} Capetti, A., Celotti, A., Chiaberge, M.,
et al.\ 2002, \aap, 383, 104
\bibitem[Cassaro et al.(1999)]{cas99} Cassaro, P., Stanghellini, C., Bondi, M.,
et al.\ 1999, \aaps, 139, 601
\bibitem[Dallacasa et al.(2000)]{dal00} Dallacasa, D., Stanghellini, C.,
Centonza, M., \& Fanti, R.\ 2000, \aap, 363, 887
\bibitem[Ebeling et al.(2002)]{ebe02} Ebeling, H., Mullis, C.~R., \& Tully,
R.~B.\ 2002, \apj, 580, 774
\bibitem[Fanaroff \& Riley(1974)]{fan74} Fanaroff, B.~L., \& Riley, J.~M.\
1974, \mnras, 167, 31P
\bibitem[Fanti et al.(1990)]{fan90} Fanti, R., Fanti, C., Schilizzi, R.~T., et
al.\ 1990, \aap, 231, 333
\bibitem[Fanti et al.(1995)]{fan95} Fanti, C., Fanti, R., Dallacasa, D., et
al.\ 1995, \aap, 302, 317
\bibitem[Fanti et al.(1986)]{fan86} Fanti, C., Fanti, R., de Ruiter, H.~R., \&
Parma, P.\ 1986, \aaps, 65, 145
\bibitem[Fanti et al.(1987)]{fan87} Fanti, C., Fanti, R., de Ruiter, H.~R., \&
Parma, P.\ 1987, \aaps, 69, 57
\bibitem[Fanti \& Fanti(2003)]{fan03} Fanti, C., \& Fanti, R.\ 2003,
Astronomical Society of the Pacific Conference Series, 300, 81
\bibitem[Feretti \& Giovannini(1985)]{fer85} Feretti, L., \& Giovannini, G.\
1985, \aap, 147, L13
\bibitem[Ghisellini \& Celotti(2001)]{ghi01} Ghisellini, G.~\& Celotti, A.\
2001, \aap, 379, L1
\bibitem[Giovannini et al.(1999)]{gio99} Giovannini, G., Taylor, G.~B.,
Arbizzani, E., et al.\ 1999, \apj, 522, 101
\bibitem[Giovannini et al.(2001)]{gio01} Giovannini, G., Cotton, W.~D.,
Feretti, L., Lara, L., \& Venturi, T.\ 2001, \apj, 552, 508
\bibitem[Giovannini et al.(2005)]{gio05} Giovannini, G., Taylor, G.~B.,
Feretti, L., et al.\ 2005, \apj, 618, 635
\bibitem[Giroletti et al.(2003)]{gir03} Giroletti, M., Giovannini, G., Taylor,
G.~B., et al.\ 2003, \aap, 399, 889
\bibitem[Giroletti et al.(2004)]{gir04} Giroletti, M., Giovannini, G., Feretti,
L., et al.\ 2004, \apj, 600, 127
\bibitem[Giroletti et al.(2005)]{gir05} Giroletti, M., Taylor, G.~B., \&
Giovannini, G.\ 2005, \apj, 622, 178
\bibitem[Gugliucci et al.(2005)]{gug05} Gugliucci, N.~E., Taylor, G.~B., Peck,
A.~B., \& Giroletti, M.\ 2005, \apj, 622, 136
\bibitem[Ho et al.(1997)]{ho97} Ho, L.~C., Filippenko, A.~V., \& Sargent,
W.~L.~W.\ 1997, \apjs, 112, 315
\bibitem[Hummel et al.(1991)]{hum91} Hummel, E., Beck, R., \& Dettmar, R.-J.\
1991, \aaps, 87, 309
\bibitem[Kaiser et al.(1997)]{kai97} Kaiser, C.~R., Dennett-Thorpe, J., \&
Alexander, P.\ 1997, \mnras, 292, 723
\bibitem[Kunert-Bajraszewska et al.(2004)]{kun04} Kunert-Bajraszewska, M.,
Marecki, A., \& Spencer, R.~E.\ 2004, European VLBI Network on New Developments
in VLBI Science and Technology, 73
\bibitem[Lara et al.(2004)]{lar04} Lara, L., Giovannini, G., Cotton, W.~D., et
al.\ 2004, \aap, 421, 899
\bibitem[Ledlow \& Owen(1996)]{led96} Ledlow, M.~J.~\& Owen, F.~N.\ 1996, \aj,
112, 9
\bibitem[Marecki et al.(2003)]{mar03} Marecki, A., Spencer, R.~E., \& Kunert,
M.\ 2003, Publications of the Astronomical Society of Australia, 20, 46
\bibitem[Marscher(1987)]{mar87} Marscher, A.~P.\ 1987, Superluminal Radio
Sources, 280
\bibitem[Morganti et al.(1997)]{mor97} Morganti, R., Parma, P., Capetti, A., et
al.\ 1997, \aaps, 126, 335
\bibitem[Morganti et al.(2003)]{mor03} Morganti, R., Oosterloo, T.~A., Capetti,
A., et al.\ 2003, \aap, 399, 511
\bibitem[Murgia \& Fanti(1996)]{mur96}Murgia, M., \& Fanti, R. 1996, Rapporto
Interno IRA, 228/96
\bibitem[Murgia et al.(1999)]{mur99} Murgia, M., Fanti, C., Fanti, R., et al.\
1999, \aap, 345, 769
\bibitem[Narayan \& Yi(1995)]{nar95} Narayan, R.~\& Yi, I.\ 1995, \apj, 452,
710
\bibitem[Ochsenbein et al.(2000)]{och00} Ochsenbein, F., Bauer, P., \& Marcout,
J.\ 2000, \aaps, 143, 23
\bibitem[O'Dea(1998)]{ode98} O'Dea, C.~P.\ 1998, \pasp, 110, 493
\bibitem[O'Dea \& Baum(1997)]{ode97} O'Dea, C.~P., \& Baum, S.~A.\ 1997, \aj,
113, 148
\bibitem[Owsianik \& Conway(1998)]{ows98} Owsianik, I., \& Conway, J.~E.\ 1998,
\aap, 337, 69
\bibitem[Parma et al.(1986)]{par86} Parma, P., de Ruiter, H.~R., Fanti, C., \&
Fanti, R.\ 1986, \aaps, 64, 135
\bibitem[Polatidis et al.(2002)]{pol02} Polatidis, A.~G., Conway, J.~E., \&
Owsianik, I.\ 2002, Proceedings of the 6th EVN Symposium, 139
\bibitem[Readhead et al.(1994)]{rea94} Readhead, A.~C.~S., Xu, W., Pearson,
T.~J., Wilkinson, P.~N., \& Polatidis, A.~G.\ 1994, Compact Extragalactic Radio
Sources, 17
\bibitem[Salvati et al.(1998)]{sal98} Salvati, M., Spada, M., \& Pacini, F.\
1998, \apjl, 495, L19
\bibitem[Sanghera et al.(1995)]{san95} Sanghera, H.~S., Saikia, D.~J., Luedke,
E., et al.\ 1995, \aap, 295, 629
\bibitem[Tinti et al.(2005)]{tin05} Tinti, S., Dallacasa, D., de Zotti, G.,
Celotti, A., \& Stanghellini, C.\ 2005, \aap, 432, 31
\bibitem[Urry \& Padovani(1995)]{urr95} Urry, C.~M., \& Padovani, P.\ 1995,
\pasp, 107, 803
\bibitem[Wilkinson et al.(1994)]{wil94} Wilkinson, P.~N., Polatidis, A.~G.,
Readhead, A.~C.~S., Xu, W., \& Pearson, T.~J.\ 1994, \apjl, 432, L87
\bibitem[Worrall \& Birkinshaw(2000)]{wor00} Worrall, D.~M., \& Birkinshaw, M.\
2000, \apj, 530, 719
\end{thebibliography}

\end{document}